\documentclass[twoside,11pt,onehalfspacing]{article}

\usepackage{fullpage}
\usepackage{authblk}
\usepackage[super,comma,compress]{natbib}
\usepackage{graphicx}
\pdfoutput=1

\usepackage{booktabs}
\usepackage{subfigure}
\usepackage{tabularx}
\usepackage{multirow}
\usepackage{xcolor}
\usepackage{tikz}
\usepackage{amsmath}

\usepackage{lipsum}
\usepackage{caption}
\usepackage{hyperref}
\usepackage{subfigure}
\usepackage{amsfonts}
\usepackage{ulem}
\usepackage{xcolor}
\usepackage{multirow}
\usepackage{graphicx}
\usepackage{amsmath}
\usepackage{amssymb}
\usepackage[small,compact]{titlesec}
\usepackage{amsmath}
\usepackage{siunitx} 
\usepackage{booktabs}
\usepackage{float}
\usepackage{accents}
\usepackage{hhline}
\usepackage{float}
\restylefloat{table}
\floatstyle{plaintop}
\restylefloat{table}
\usepackage{tabularx}
\usepackage[section]{placeins}
\usepackage[ruled, noline, noend, linesnumbered]{algorithm2e}

\newcommand{\doseVecOpt}{\hat{\textbf{d}}_{p}}
\newcommand{\groundTruth}{\hat{\textbf{s}}_{p}}
\newcommand{\submission}{\textbf{s}_{p}}
\newcommand{\privateData}{\mathcal{P}}
\newcommand{\openkbpData}{\mathcal{O}}
\newcommand{\Praw}{\privateData^\text{raw}}
\newcommand{\Pclean}{\privateData^\text{clean}}
\newcommand{\Oraw}{\openkbpData^\text{raw}}
\newcommand{\Oclean}{\openkbpData^\text{clean}}
\newcommand{\Oaug}{\openkbpData^\text{aug}}
\newcommand{\Otrain}{\openkbpData^\text{train}}
\newcommand{\Oval}{\openkbpData^\text{val}}
\newcommand{\Otest}{\openkbpData^\text{test}}

\newcommand{\AllRaw}{\Praw\cup\Oraw}
\newcommand{\holdOutSet}{\mathcal{O}^h}

\newcommand{\D}[2]{\text{D}_\text{#1}^{{#2}}}  

\newcommand{\Devals}[3]{\D{#1}{#2}({#3})}

\newcommand{\Dupperbound}[2]{\overline{\text{D}}_\text{#1}^\text{#2}}
\newcommand{\Dlowerbound}[2]{\underline{\text{D}}_\text{#1}^\text{#2}}
\newcommand{\Dups}[3]{\Devals{#1}{#2}{#3}/\Dupperbound{#1}{#2}}
\newcommand{\Ddowns}[3]{\Dlowerbound{#1}{#2}/\Devals{#1}{#2}{#3}}

\newcommand{\plans}{p\in\mathcal{P}}
\newcommand{\OpenKBPplans}{\Oval\cup\Otest}
\newcommand{\oars}{\mathcal{I}_p}
\newcommand{\targets}{\mathcal{T}_p}
\newcommand{\rois}{r\in\mathcal{I}_p\cup\mathcal{T}_p}
\newcommand{\metrics}{c\in\mathcal{C}_r}
\newcommand{\noise}[2]{\varepsilon\sim \mathcal{U}({#1}, {#2})}
\newcommand{\doseError}{\alpha_p}
\newcommand{\doseScore}{A_h}
\newcommand{\dvhError}{\beta^r_{p, c}}
\newcommand{\dvhScore}{B_h}
\def\ab#1{\color{black}#1}
\makeatletter
\renewcommand\NAT@citesuper[3]{\ifNAT@swa
\if*#2*\else#2\NAT@spacechar\fi
\unskip\kern\p@\textsuperscript{\NAT@@open#1\if*#3*\else,\NAT@spacechar#3\fi\NAT@@close}%
   \else #1\fi\endgroup}
\makeatother
\begin{document}

\title{OpenKBP: The open-access knowledge-based \\planning grand challenge}

  \renewcommand*{\Authands}{, }
    \author[1]{Aaron Babier}
  \author[1]{Binghao Zhang}
  \author[1]{Rafid Mahmood}
  \author[2]{Kevin L. Moore}
  \author[3]{Thomas G. Purdie}
  \author[3]{Andrea L. McNiven}
  \author[1]{Timothy C. Y. Chan\vspace*{-0.25cm}}
  \affil[1]{Department of Mechanical \& Industrial Engineering, University of Toronto}
  \affil[2]{Department of Radiation Oncology, University of California, San Diego}
  \affil[3]{Radiation Medicine Program, Princess Margaret Cancer Centre}

  \date{}
\maketitle

\begin{abstract}

The purpose of this work is to advance fair and consistent comparisons of dose prediction methods for knowledge-based planning (KBP) in radiation therapy research. We hosted OpenKBP, a 2020 AAPM Grand Challenge, and challenged participants to develop the best method for predicting the dose of contoured CT images. The models were evaluated according to two separate scores: (1) dose score, which evaluates the full 3D dose distributions, and (2) dose-volume histogram (DVH) score, which evaluates a set DVH metrics. Participants were given the data of 340 patients who were treated for head-and-neck cancer with radiation therapy. The data was partitioned into training ($n=200$), validation ($n=40$), and testing ($n=100$) datasets. All participants performed training and validation with the corresponding datasets during the validation phase of the Challenge, and we ranked the models in the testing phase based on out-of-sample performance. The Challenge attracted 195 participants from 28 countries, and 73 of those participants formed 44 teams in the validation phase, which received a total of 1750 submissions. The testing phase garnered submissions from 28 teams. On average, over the course of the validation phase, participants improved the dose and DVH scores of their models by a factor of 2.7 and 5.7, respectively. In the testing phase one model achieved significantly better dose and DVH score than the runner-up models. Lastly, many of the top performing teams reported using generalizable techniques (e.g., ensembles) to achieve higher performance than their competition. This is the first competition for knowledge-based planning research, and it helped launch the first platform for comparing KBP prediction methods fairly and consistently. The OpenKBP datasets are available publicly to help benchmark future KBP research, which has also democratized KBP research by making it accessible to everyone.

\end{abstract}

\section{Introduction}

\begin{figure}[H]
\centering
         \includegraphics[width=1\linewidth]{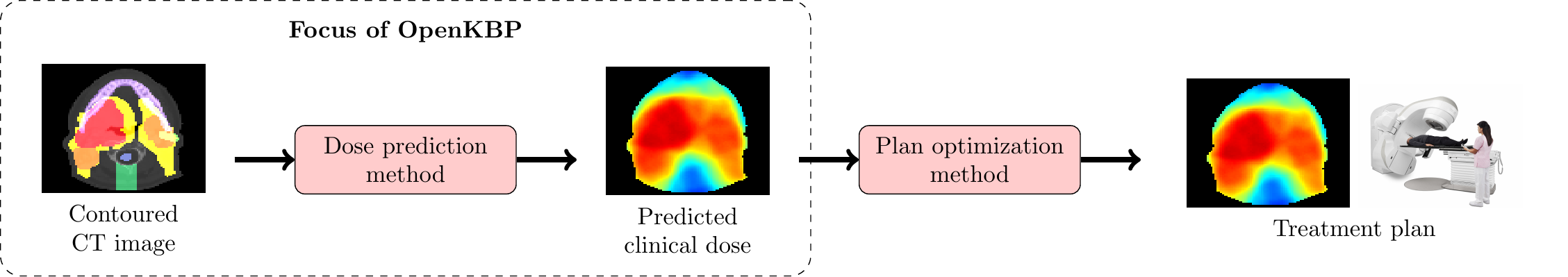}
        \caption{Overview of a complete knowledge-based planning pipeline.}
        \label{fig:automated-kbp}
\end{figure}

The increasing demand for radiation therapy to treat cancer has led to a growing focus on improving patient flow in clinics.\cite{Atun:2015aa} Knowledge-based planning (KBP) methods promise to reduce treatment lead time by automatically generating patient-specific treatment plans, thereby streamlining the treatment planning process.\cite{Sharpe:2014aa} KBP methods are generally formulated as two-stage pipelines (see Figure~\ref{fig:automated-kbp}). In most cases, the first stage is a machine learning (ML) method that predicts the dose distribution that should be delivered to a patient based on contoured CT images, and the second stage is an optimization model that generates a treatment plan based on the predicted dose distribution.\cite{McIntosh:2017aa,Babier:2020ab} 

Research into dose prediction has experienced major growth in the past decade,\cite{Ge:2019aa} 
in part due to the growing sophistication of machine learning and optimization methods in conjunction with advances in computational technology. There are two main branches of dose prediction methods: (1) those that predict summary statistics (e.g., dose-volume features)\cite{PCAYuan:2012aa,PCAZhu:2011aa,appenzoller:2012predicting,Babier:2018b} and (2) those that predict entire 3D dose distributions.\cite{McIntosh:2016aa,Shiraishi:2016aa,Kearney:2018aa,Nguyen:2019aa,GANCER,Babier:2020aa} Both branches of dose prediction methods use a wide range of methodologies, e.g., linear regression,\cite{Babier:2018b} principal component analysis,\cite{PCAYuan:2012aa,PCAZhu:2011aa} random forest,\cite{McIntosh:2016aa} neural networks.\cite{Shiraishi:2016aa,Kearney:2018aa,Nguyen:2019aa,GANCER,Babier:2020aa} All of this KBP research is performed in close collaboration with radiation therapy clinics using private clinical datasets that are generated via local planning protocols.\cite{Ge:2019aa}


Development of KBP models is further challenged by the lack of large open-access datasets and standardized evaluation metrics. 
Existing open-access radiation therapy datasets cater to optimization\cite{Craft:2014aa,Breedveld:2017aa} or classification problems (e.g., segmentation, prognosis).\cite{Clark:2013aa} Researchers that develop dose prediction models must rely on their own private clinical datasets and different evaluation metrics, which makes it difficult to objectively and rigorously compare the quality of different prediction approaches at a meaningful scale.\cite{Ge:2019aa} \ab{As a result, researchers must recreate published dose prediction models to benchmark their new models via a common dataset and set of evaluation metrics}.\cite{Babier:2020ab,McIntosh:2017ab} In contrast, open-access datasets and standardized metrics are staples of thriving artificial intelligence-driven fields, as evidenced by the uptake of the CIFAR\cite{cifar} and ImageNet\cite{imageNet} datasets in the computer vision community over the past decade.

We launched the Open Knowledge-Based Planning (OpenKBP) Grand Challenge to advance knowledge-based planning by 1) providing a platform to enable fair and consistent comparisons of dose prediction methods and 2) developing the first open-access dataset for KBP. Participants of the Challenge used the dataset to train, test, and compare their prediction methods, using a set of standardized evaluation metrics. The data and accompanying code-base is freely available at~\url{https://github.com/ababier/open-kbp} for KBP researchers to use going forward. 

\section{Methods and Materials}
We first describe our process for building and validating the dataset for the Challenge. We then describe how the Challenge was organized and delivered. Finally, we provide an analysis of the Challenge results. This study was approved by the Research Ethics Board at the University of Toronto. 

\subsection{Data Processing}
\begin{figure}[H]
\centering
         \includegraphics[width=1\linewidth]{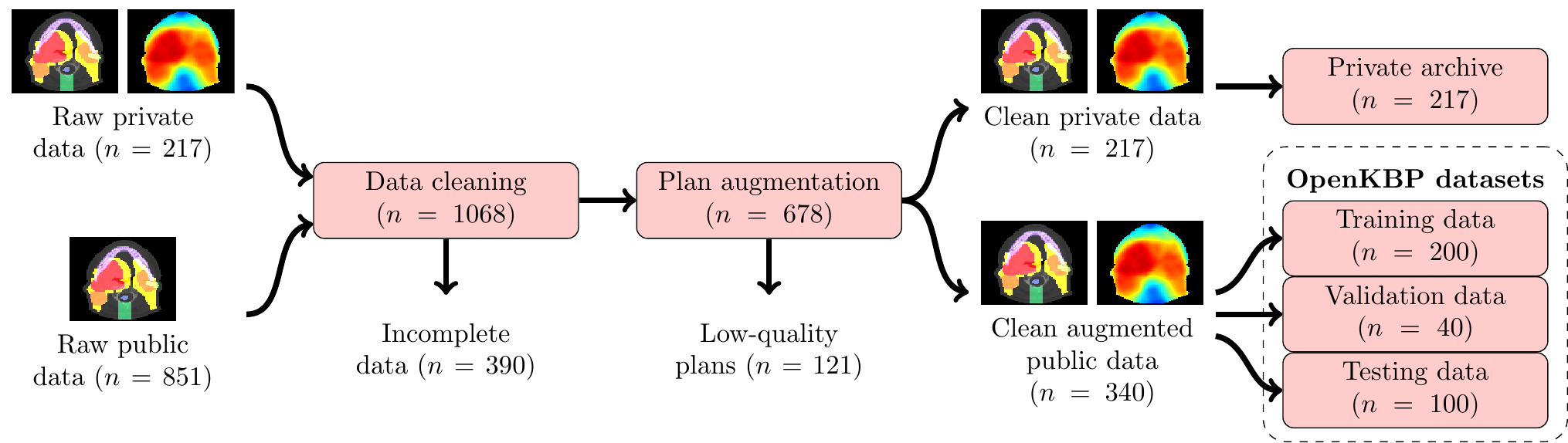}
        \caption{Overview of the data processing pipeline. $n$ represents the number of patients at each stage of the pipeline.}
        \label{fig:data-processing}
\end{figure}

Figure~\ref{fig:data-processing} depicts our data processing approach at a high level, which consisted of four steps: (i) data acquisition, (ii) data cleaning, (iii) plan augmentation, and (iv) data partitioning.

\subsubsection{Acquiring the raw data}
We obtained the Digital Imaging and Communications in Medicine (DICOM) files of 217 patients, which we call the raw private data (denoted by $\Praw$), who were treated for oropharyngeal cancer at Princess Margaret Cancer Center. Each file included a treatment plan that was delivered from nine approximately equispaced coplanar fields with 6 MV, step-and-shoot, intensity-modulated radiation therapy (IMRT). Each patient was prescribed 70 Gy in 35 fractions, with 70 Gy to the high-dose planning target volume (PTV70), 63 Gy to the mid-dose planning target volume (PTV63), and 56 Gy to the low-dose planning target volume (PTV56); a PTV63 was only contoured in 130 of the patients. All plans included CT images, contours for regions-of-interest (ROIs), and the dose distributions based on a consistent set of planning protocols. 


We also retrieved clinical DICOM data for 851 patients, which we call the raw public data (denoted by $\Oraw$), from four public data sources \cite{data-HN-Cetuximab,data-HN-Pet-CT,data-HNSCC,data-TCGA} hosted on The Cancer Imaging Archive (TCIA).\cite{Clark:2013aa} The data was originally sourced from twelve different institutions between 1999 and 2014. Each file contained CT images and contours for the regions of interest (ROIs). This collection of files contained several inconsistencies because the data originated from different institutions. For example, different institutions may have employed different dose levels, fractionation schemes, ROI naming conventions, languages (English versus French nomenclature), PTV margins (isotropic versus anisotropic margins from CTV), and treatment modalities (3D conformal radiation therapy (3DCRT) versus IMRT). 


\subsubsection{Data cleaning}
In order to standardize and improve the homogeneity of the datasets, we employed a sequence of data cleaning procedures. First, we relabeled all of ROIs according to a consistent nomenclature. For each patient $p\in\AllRaw$, we included organ-at-risk (OAR) contours for the brainstem, spinal cord, right parotid, left parotid, larynx, esophagus, and mandible; let $\oars$ denote this set of OARs for a patient $p$. All other OAR contours were deleted. Also, an OAR was omitted from $\oars$ if it was not contoured in the clinical plan (e.g., a patient whose left parotid was not contoured would not have it in the set $\oars$). To construct the set of targets $\targets$, we identified the low-, mid-, and high-dose targets based on their relative dose levels\cite{Mayo:2018aa} and relabeled them as PTV56, PTV63, and PTV70, respectively. 
Any region with overlapping PTVs was relabeled as a single PTV with a dose-level equal to that of the highest dose-level of those overlapping PTVs. 

Next, we modified target contours in the raw public dataset ($\Oraw$) to match the protocols from the private dataset ($\Praw$). These modifications helped to fix some of the inconsistencies in contouring (e.g., no PTV margins, anisotropic PTV margins) that were present in the raw public dataset. Every PTV was expanded to include the voxels within $5~\text{mm}$ of its respective clinical target volume (CTV); the PTV was left unchanged in cases where there was no CTV contour associated with the PTV. Every PTV was also clipped to be no closer than $5~\text{mm}$ from the surface of the patient. 

We generated dose influence matrices for each patient in the public dataset $\Oraw$ based on 6 MV step-and-shoot IMRT with nine equispaced coplanar fields at $0^\circ$, $40^\circ$, $\hdots$, $320^\circ$. Those fields were divided into a set of beamlets $\mathcal{B}$ that were each $5~\text{mm}\times5~\text{mm}$. Every patient was also divided into a set of voxels $\mathcal{V}^p$ that were downsampled to fill axial slices of dimension $128\times128$. The relationship between the intensity $w_b$ of beamlet $b$ and dose $d_v$ deposited to voxel $v$ was calculated in \textsf{MATLAB} using the \textsf{IMRPT} library in \textsf{A Computational Environment for Radiotherapy Research},\cite{CERR} which we used to form the elements $D_{v, b}$ of each patient's dose influence matrix. The dose to a voxel $v$ was calculated as:

\begin{equation*}
\label{IM}
d_v= \sum\limits_{b \in \mathcal{B}} D_{v,b}w_b,\ \forall v \in \mathcal{V}^p, \ \forall \plans.
\end{equation*}

To prepare the patient data for deep learning models, we framed each patient in a $128\times128\times128$ voxel tensor in two steps. First, we calculated the weighted average position of each patient $p$, using $\sum\limits_{b \in \mathcal{B}} D_{v,b}$ as the weight for each voxel $v \in\mathcal{V}^p$. Second, we applied a bounding box centered on that weighted average position with dimensions of $128\times128\times128$ voxels. We added zero-padding where necessary to ensure consistent tensor volumes. Over the course of the data cleaning phase, 390 patients were removed from the public dataset ($\Oraw$) for a variety of reasons including missing target contours and issues generating a valid dose influence matrix. No patients were removed from our private dataset. At the end of the data cleaning step, we had clean private $\Pclean$ and public $\Oclean$ datasets consisting of 217 and 461 patients, respectively.

\subsubsection{Plan augmentation}
\begin{figure}[H]
\centering
         \includegraphics[width=1\linewidth]{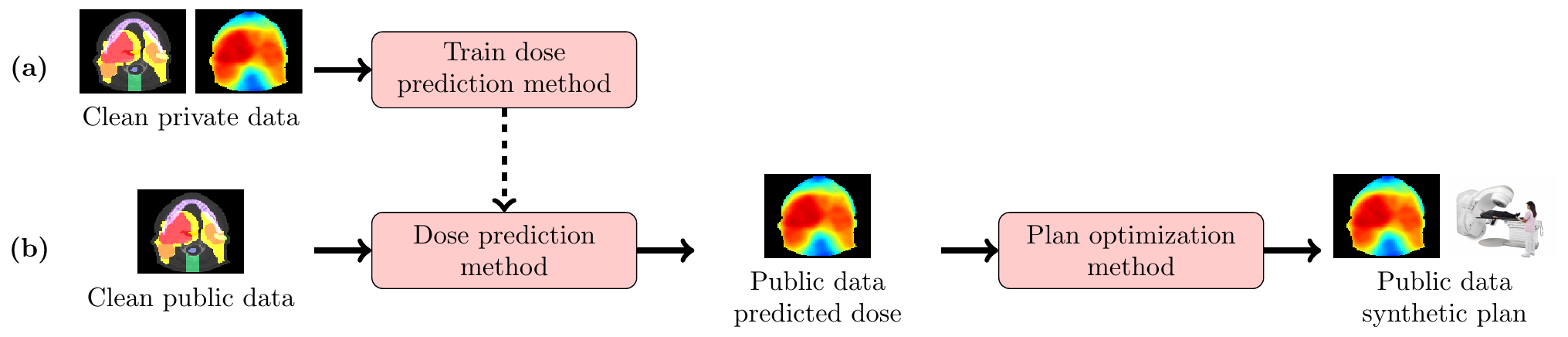}
        \caption{Overview of the plan augmentation process, which is a two phased approach: (a) the clean private clinical dataset is used to train the dose prediction method, and then (b) the trained method is used in a complete KBP pipeline that intakes the cleaned public data to generate synthetic plans. }
        \label{fig:plan-augmentation}
\end{figure}

Next, we generated synthetic plans for each patient in the clean public dataset and only retained the associated dose distribution. These synthetic plans were generated using a variation of a published automated KBP pipeline, \cite{Babier:2020aa} which was trained using the cleaned clinical plans from our private dataset $\Pclean$. Figure~\ref{fig:plan-augmentation} illustrates the plan augmentation process.

The dose prediction model in the KBP pipeline was a conditional generative adversarial network (GAN)\cite{isola:2017image} with the same architecture as the 3D-GAN in Babier \textit{et al.} 2020.\cite{Babier:2020aa} It uses two neural networks: (1) a generator that predicts the dose distribution for a contoured CT image; and (2) a discriminator that predicts whether the input is a predicted or clinical dose distribution. We trained the generator to minimize the mean absolute difference between the predicted and clinical dose distributions, which we regularized with the discriminator to encourage the generator to make predictions resembling clinical dose distributions. Between each batch update of the generator, we also trained the discriminator to minimize a binary-cross-entropy loss function. This GAN model was trained for 200 epochs using the clean private dataset of 217 treatment plans, and it was implemented in \textsf{Tensorflow 1.12.3} on a Nvidia~$1080$Ti graphic processing unit (GPU) with $12~\text{GB}$ of video memory. 

As part of the plan optimization, we added seven optimization structures to each patient in the public dataset to encourage high-quality synthetic plans. \ab{All of these optimization structures are based on structures that were used to optimize the plans in our private clinical dataset.} These structures were not included in the final Challenge datasets. The first optimization structure was called limPostNeck, which is used to limit dose to the posterior neck. The limPostNeck includes all of the non-target voxels between the posterior aspect of a $3~\text{mm}$ expansion of the spinal cord and the patient posterior; there were $12$ cases where no spinal cord was contoured where we extended the brainstem inferiorly to approximate the spinal cord to make the limPostNeck. All spinal cord and target voxels were removed from the limPostNeck. The other six optimization structures were PTV rings, which we added to encourage high dose gradients around the PTVs. We used $2~\text{mm}$ and $6~\text{mm}$ rings that include voxels within $0~\text{mm}$ to $2~\text{mm}$ and $2~\text{mm}$ to $6~\text{mm}$ of the PTV, respectively. Any overlap between rings was eliminated by removing voxels in those overlapping regions from the ring of PTV with the lower dose-level. All target voxels were also removed from the rings. 

The plan optimization method was a two-stage approach to inverse planning.\cite{Babier:2018a} In the first stage, we estimate the objective function weights for a conventional inverse planning model that makes a predicted dose distribution optimal. In the second stage, we use the estimated weights and solve the inverse planning model to generate a synthetic treatment plan. The objective of the planning model was to minimize the sum of 114 functions: seven per OAR, three per target, and seven per optimization structure. The functions for each OAR evaluated the mean dose; maximum dose; and average dose above 0.25, 0.50, 0.75, 0.90, and 0.975 of the maximum predicted dose to that OAR. The functions for each target evaluated the mean dose, maximum dose, average dose below the target dose level, and average dose 5\% above the target dose level (e.g., average dose above 73.5 Gy in the PTV70). The functions for each optimization structure were the same as the OAR functions. To ensure that all plans had a similar degree of fluence complexity, all synthetic plans were constrained to a sum-of-positive-gradients (SPG) value of 65.\cite{Craft:2007aa} Both optimization models were solved in \textsf{Python 3.6} using \textsf{Gurobi 9.0.1} (Gurobi Optimization, TX, US) to generate a dose distribution $\doseVecOpt$ for each patient in the clean public dataset. 

We used Algorithm~\ref{algo:filter} to correct or remove low-quality plans that were generated by our plan augmentation process (i.e., the process in Figure~\ref{fig:plan-augmentation}). The algorithm curated a set of patients $\Oaug$ with high-quality dose distributions $\groundTruth$, which were based on the dose distributions $\doseVecOpt$ for the plans of patients in the clean public dataset $\Oclean$. The algorithm retained any patients that had plans with a high-dose target that received a higher mean dose or 1$^\text{st}$ percentile dose ($\D{99}{r}$) than the mid-dose or low-dose targets (line 3). The entire dose was then multiplicatively scaled so that maximum dose to the high-dose target $\Devals{max}{\text{PTV70}}{\groundTruth}$ was no lower than the lowest maximum dose ($\Dlowerbound{max}{\text{PTV70}}$) observed in the plans of the patients from our clean private dataset $\Pclean$ (line 4). For each instance where we scaled dose by a constant factor, we also scaled the dose by a random factor $\varepsilon$ that was sampled from a uniform distribution between 1.00 and 1.05 (i.e., $\noise{1.00}{1.05}$) for two reason: (1) we did not want participants to learn a strict cutoff and (2) strict cutoffs are not realistic. Next, we reduced the dose $\groundTruth$ so that, for each ROI $\rois$, it delivered a maximum dose $\D{max}{r}$, mean dose $\D{mean}{r}$, and dose to 99\% of voxels $\D{99}{r}$ that was lower than any plan in our private clinical dataset (line 7). We denote the highest value observed in the clinical plans with a bar (e.g., $\Dupperbound{c}{r}$ for a criteria $c$ and ROI $r$). Lastly, a patient $p$ was added to $\Oaug$ if that patient's respective dose $\groundTruth$ had a maximum dose to the high-dose target that was between the lower and upper bounds that we observed in our private set of clinical plans (line 8). The final size of $\Oaug$ was 340.


\begin{algorithm}[H]
\DontPrintSemicolon
 $\Oaug \gets \{\}$
 
 \For{$p\in\Oclean$}{
  	\quad $\groundTruth \gets \doseVecOpt$
  	
 	\quad \If{\upshape$\Devals{mean}{PTV70}{\groundTruth} \ge \Devals{mean}{t}{\groundTruth}$  \ or $\Devals{99}{PTV70}{\groundTruth} \ge \Devals{99}{t}{\groundTruth},\ \forall t\in\mathcal{T}_p$}{
	\qquad $\groundTruth \gets \groundTruth \times \text{max}(1,\ \Ddowns{max}{PTV70}{\groundTruth} \times\noise{1.00}{1.05})$
	
 	\qquad \For{$\rois$}{
	 	\quad\qquad \For{$c\in\{\text{\upshape max, mean, 99}\}$}{
		\qquad\qquad $\groundTruth \gets \groundTruth \times \text{min}(1,\ \Dups{c}{r}{\groundTruth} \times\noise{0.97}{1.00})$
		}
 	}
	\qquad \If{\upshape$\Dlowerbound{max}{PTV70} > \Devals{max}{PTV70}{\groundTruth} > \Dupperbound{max}{PTV70}$}{
		\quad\qquad $\Oaug \gets \Oaug \cup \{p\}$ 
		}
}
}
 \caption{Improve low-quality plans where possible and construct the set of public patients with high-quality synthetic plan dose distributions $\groundTruth$.}
 \label{algo:filter}
\end{algorithm}

\subsubsection{Validation of final competition datasets}
We evaluated the distribution of synthetic dose $\groundTruth$ quality over every patient $p\in\Oaug$ by comparing it to the distribution of the clinical dose quality over every patient $p\in\Pclean$.  We measured quality using the set of DVH criteria used in the Challenge. The distribution of DVH criteria over the population of synthetic doses and clinical doses was visualized with a box plot for each set of criteria.
For each of the DVH criteria, we used a one-sided Mann-Whitney $U$ test to determine whether the synthetic doses were inferior (null hypothesis) or non-inferior (alternative hypothesis) to the clinical doses, based on an equivalence interval of 2.1 Gy (i.e., 3\% of the high-dose level).\cite{Althunian:2017aa} Lower values were better for $\D{mean}{}$, $\D{0.1cc}{}$, and $\D{1}{}$; higher values were better for $\D{95}{}$ and $\D{99}{}$. For these and all subsequent hypothesis tests, $P < 0.05$ was considered significant. 

The final public dataset $\Oaug$ was randomly split into training $\Otrain$, validation $\Oval$, and testing $\Otest$ datasets with 200, 40, and 100 patients, respectively. Every patient in these datasets had a synthetic dose distribution ($\groundTruth$), CT images, structure masks, feasible dose mask (i.e., voxels $v\in\mathcal{V}^p$ such that $\sum\limits_{b \in \mathcal{B}} D_{v,b} > 0$), and voxel dimensions. This data was released to the participants in phases as described in the next section. A detailed description of the data format and files is given in Appendix A.

\subsection{Challenge Description}
OpenKBP was hosted as an online competition using \textsf{CodaLab} (Microsoft Research, Redmond, WA). Participants could compete in the Challenge as a member of a team or as individuals (i.e., a team of one). The Challenge proceeded in two phases. In the first (validation) phase, teams developed their models and compared their performance in real time to other teams via a public leaderboard. In the second (testing) phase, teams submitted their dose predictions for a new unseen dataset, and we compared their performance to other teams via a hidden leaderboard to determine the final rankings for the Challenge.

\subsubsection{Challenge timeline}
The Challenge took place over four months in 2020. Individuals could register to participate in the Challenge anytime after it started on February 21, 2020, which is also when the training and validation data was released to start the first (validation) phase of the Challenge. Three months later, on May 22, 2020, the testing data was released to start the second (testing) phase of the Challenge, which ended ten days later on June 1, 2020. The final competition rankings (based on testing phase performance) were released four days later on June 5, 2020. The Challenge also coincided with the beginning of the COVID-19 pandemic.\cite{Dong:2020aa} As a result, we extended the validation phase to accommodate for the challenges posed by the pandemic. The result was about a one-month delay compared to the originally planned timeline. 

\subsubsection{Participants}
OpenKBP was designed with a view towards having a low barrier to entry. Registration was free and open to anyone. We also offered comprehensive instructions to set up free, high-quality compute resources via \textsf{Google Colab}, for those teams who did not have access to sufficient computational resources otherwise.\cite{Carneiro:2018aa}

To understand the make-up of the OpenKBP community, we collected demographic information from every participant via a two-part registration survey (see Appendix B). The first part of the survey, which was mandatory, collected professional information including their past KBP research experience, primary research area, and academic/industry affiliations. The second part of the survey, which was optional, collected equity, diversity, and inclusion (EDI) data including how participants self-identify in terms of gender, race, and disability status, using terminology from the United States Census Bureau.

\subsubsection{Evaluation metrics}
Teams predicted dose distributions or dose-volume histograms for a set of patients and submitted those predictions to our competition on \textsf{CodaLab}. For each patient $p$, we compared the submitted prediction $\submission$ to the corresponding synthetic plan dose distribution $\groundTruth$ via two error measures (1) \emph{dose error}, $\doseError$, which measures the mean absolute difference between a submission and its corresponding synthetic plan (\ab{i.e., mean absolute voxel-by-voxel difference in dose)}, and (2) \emph{DVH error}, $\dvhError$, which measures the absolute difference in DVH criteria between a submission and its corresponding synthetic plan. The dose error $\doseError$ was chosen as a general measure of prediction quality that is not radiation therapy specific. It was only used to evaluate dose distributions (i.e., not DVH submissions), and it is calculated as 

\begin{equation}
\label{doseError}
\doseError = \frac{\left|\left| \submission - \groundTruth \right|\right|_1}{\left|\mathcal{V}^p\right|},\ \forall\ p\in\OpenKBPplans.
\end{equation}

The DVH error $\dvhError$ was chosen as a clinical measure of prediction quality that is radiation therapy specific. It involves a set of DVH criteria $\mathcal{C}_i$ and $\mathcal{C}_t$ for each OAR $i\in\mathcal{I}_p$ and target $t\in\mathcal{T}_p$, respectively. There were two OAR DVH criteria: $\D{mean}{i}$, which is the mean dose received by OAR $i$; and $\D{0.1cc}{i}$, which is the maximum dose received by 0.1cc of OAR $i$. There were also three target DVH criteria: $\D{1}{t}$, $\D{95}{t}$, and $\D{99}{t}$, which are the doses received by 1\% ($99^\text{th}$ percentile), 95\% ($5^\text{th}$ percentile), and 99\% ($1^\text{st}$ percentile) of voxels in target $t$, respectively. The DVH error was used to evaluate both dose distributions and DVHs, and it is calculated as

\begin{equation}
\label{dvhError}
\dvhError = \left| \D{c}{r}(\submission) - \D{c}{r}(\groundTruth) \right|,\ \forall\ \metrics,\ \forall\ \rois,\ \forall\ p\in\OpenKBPplans.
\end{equation}

We chose to make both error metrics absolute differences to reward models that learn to make realistic dosimetric trade-offs. This is critical because prediction models that make unrealistic trade-offs (e.g., predict low dose to an OAR that is unachievable) generally perform worse than models that make realistic trade-offs in full KBP pipelines.\cite{Babier:2018b}

Building on these error metrics, we scored submissions using \emph{dose score} $\doseScore$ and \emph{DVH score} $\dvhScore$. Both scores are a variation of mean absolute error (MAE). The dose score is the mean dose error over all patients in a hold-out set  (i.e., $\Oval$ or $\Otest$):

\begin{equation}
\label{doseScore}
\doseScore = \frac{1}{\left|\holdOutSet\right|}\sum\limits_{p\in\holdOutSet}\doseError,\ \forall\ h\in\{\text{val, test}\}.
\end{equation}

The DVH score is the mean DVH error over all criteria from the patients in a hold-out set:

\begin{equation}
\label{DVHScore}
\dvhScore = \frac{1}{\sum\limits_{p\in\holdOutSet}\sum\limits_{\rois}|\mathcal{C}_r|}\sum\limits_{p\in\holdOutSet}\sum\limits_{\rois}\sum\limits_{\metrics}\dvhError,\ \forall\ h\in\{\text{val, test}\}.
\end{equation}

Using those scores, we ranked all of the submissions to the Challenge in two streams: (1) the dose stream where the team with the lowest (i.e., best) dose score won, and (2) the DVH stream where the team with the lowest (i.e., best) DVH score won.

\subsubsection{Validation phase}
At the start of the validation phase, the full training dataset $\Otrain$ of 200 patients was released, and the teams used that data to train their models. An out-of-sample validation dataset $\Oval$, which included data for 40 patients without synthetic plan dose, was also released for teams to validate the out-of-sample performance of their models. Predictions made on the validation  dataset were submitted directly to our competition on \textsf{CodaLab} where they were scored in the cloud using the held-back synthetic plan dose. The resulting scores populated a public leaderboard, but they were not used to determine the winners of the Challenge.

\subsubsection{Testing phase}
The testing dataset $\Otest$, which included data for 100 patients without synthetic plan dose, was released at the start of the testing phase. Teams used the models they developed during the validation phase and made predictions on this new unseen testing dataset. Similar to the validation phase, all predictions were submitted to our competition on \textsf{CodaLab} where they were scored in the cloud using the held-back synthetic plan dose. However, the resulting dose and DVH scores populated the testing leaderboard that we kept hidden until the competition finished. The team that performed best on the testing leaderboard with respect to the dose and DVH score was the winner of the dose and DVH stream, respectively. Teams that submitted to the testing leaderboard also responded to a model survey (see Appendix B) to summarize their models. 

\subsection{Analysis of Challenge Outcomes}
We conducted four analyses. First, we summarized the demographics of the participants. Second, we evaluated the aggregate improvements made by the teams over the course of the validation phase. Third, we compiled and analyzed the final results from the testing phase. Fourth, we summarized common modeling techniques that were employed by the participants.

\subsubsection{Participant information}
We examined the registration information of all participants and calculated summary statistics for primary research area, past KBP research experience, country of work/study, and EDI data. We compared our aggregated EDI data to comparable data for the population of people who are employed in science and engineering (S\&E) in the United States (US)\cite{nsf_stats} and the general US population.\cite{us-population}

\subsubsection{Performance over validation phase}
\ab{As a retrospective analysis,} we evaluated the aggregate improvement of all teams over the validation phase \ab{to measure their progress throughout the Challenge}. \ab{We plotted} the dose and DVH score against a relative measure of progress towards their final model, which we call the \emph{normalized submission count} (NSC). The NSC is equal to the cumulative number of submissions a team made up to a certain point in time divided by the total number of submissions made in the validation phase. For example, if a team made 100 total submissions, the score at NSC = 0.5 represents that team's best recorded performance after their 50$^\text{th}$ submission. For each team, we recorded their best cumulative dose and DVH score in increments of 0.05 NSC. At each increment we plotted the average and the 95\% confidence interval of those scores over all teams that made more than 20 total submissions.  


\subsubsection{Final results in testing phase}
We used a one-sided Wilcoxon signed-rank test to determine whether the set of predictions of the best team in each stream had the same (null hypothesis) or lower (alternative hypothesis) error (i.e., $\doseError$ and $\dvhError$) than each set of predictions submitted by the other teams. 
To visualize the range of expected error differences, we plotted the difference in dose error over all patients ($n=100$) and the difference in DVH error over all DVH criteria ($n=1783$) between the winning submission and the runner-up submissions, for the dose stream and DVH stream, respectively.

As a retrospective sensitivity analysis, we evaluated the submissions according to an \emph{alternative} scoring function with squared error terms (i.e., $\doseError^2$ and ${\dvhError}^2$) instead of absolute error terms (i.e., $\doseError$ and $\dvhError$) to determine if the final competition standings would have changed. We refer to the competition and alternative scores as MAE-based and mean squared error (MSE)-based, respectively. As a quantitative measure of the alignment between the two ranking methods, we evaluated the rank-order correlation between the rankings for the MAE-based and MSE-based scores via Spearman's rank test.

\subsubsection{Common modeling decisions}
Finally, we present a summary of the model survey information that teams submitted during the testing phase. We present common modeling choices (e.g., model architectures), hardware, and software that teams used. Lastly, we present a set of techniques that we believe are generalizable to most dose prediction frameworks, based on what teams commonly employed. 

\section{Results}

\subsubsection{Validation of final competition datasets}
\begin{figure}[H]
\centering
\subfigure[\ OAR $\D{mean}{}$ values]{
\includegraphics[width=0.4675\linewidth]{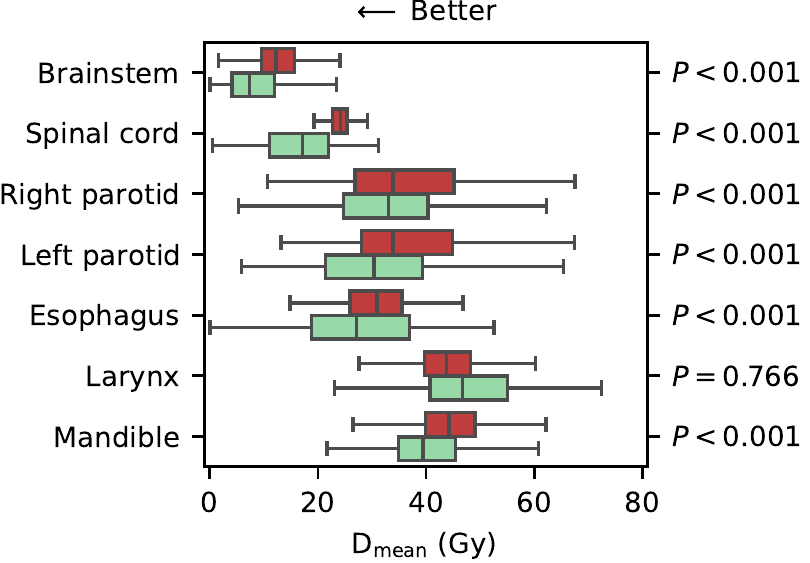} }
\subfigure[\ OAR $\D{0.1cc}{}$ values]{
\includegraphics[width=0.4675\linewidth]{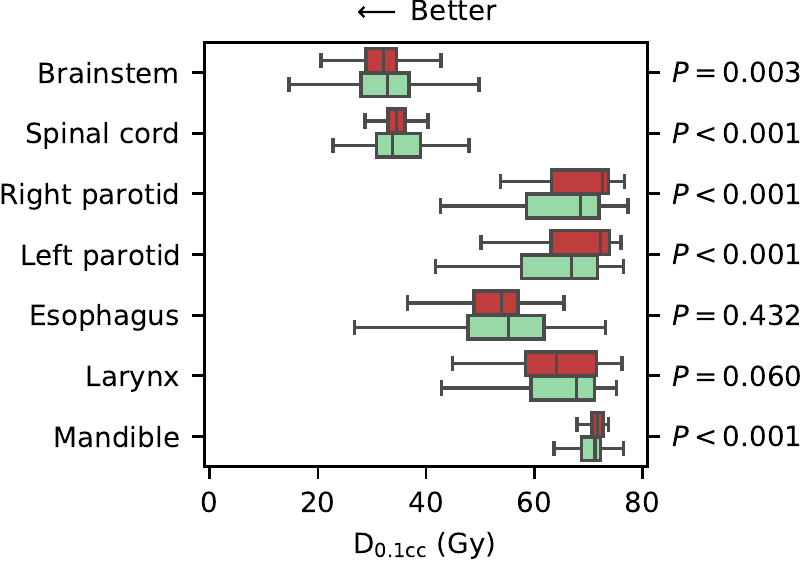} }\\
\subfigure[\ Target $\D{1}{}$ values]{
\includegraphics[width=0.31\linewidth]{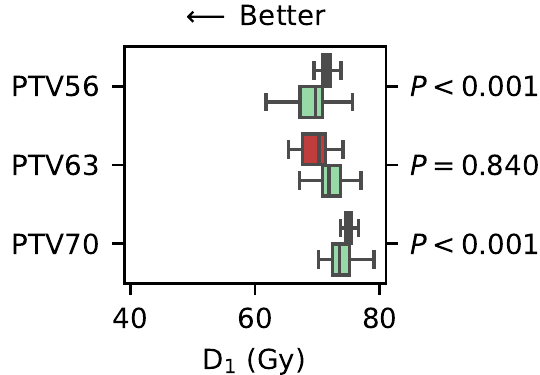} }
\subfigure[\ Target $\D{95}{}$ values]{
\includegraphics[width=0.31\linewidth]{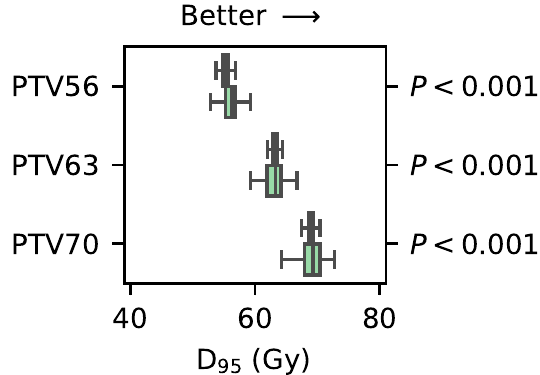}}
\subfigure[\ Target $\D{99}{}$ values]{
\includegraphics[width=0.31\linewidth]{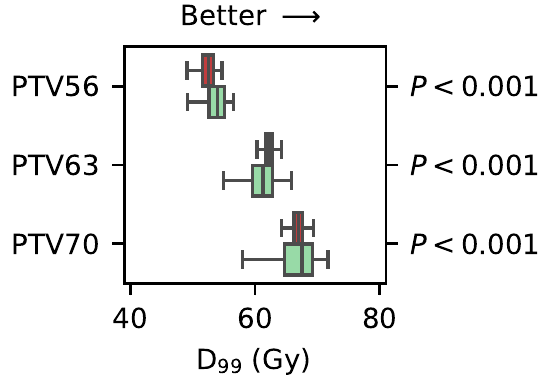}}
\includegraphics[width = 0.55\linewidth]{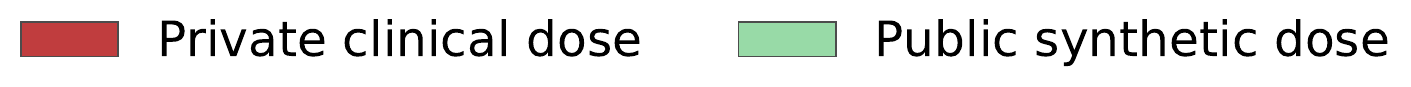}
\caption{The distribution of DVH criteria from the private clinical dose and the public synthetic dose is plotted, and the corresponding $P$-values for each criterion are on the right axes. 
The boxes indicate median and interquartile range (IQR). Whiskers extend to the minimum of 1.5 times the IQR and the most extreme outlier.}
\label{figure:plan-quality}
\end{figure}

Figure~\ref{figure:plan-quality} compares the quality of the public synthetic dose distributions to the private clinical dose distributions. The box plots in the top and bottom row summarize the performance across OAR and target DVH criteria, respectively. The public synthetic doses were non-inferior ($P<0.05$) to the clinical doses on 19 of the 23 criteria. For the remaining four criteria, the synthetic dose was 2.1 Gy worse on average than the clinical dose (3.7\% average relative difference). While the synthetic doses were not a perfect replication of the clinical doses, they are sufficiently close to representing clinical dose distributions for the purpose of this Challenge and future research that leverages this dataset.

\subsubsection{Participant information}
Table~\ref{tab:participant-info} summarizes the participation in each phase of the Challenge. Overall, 195 people registered to participate, and 73 participants were active during the validation phase. A total of 1750 submissions were made to the validation phase, which is an average of 40 submission per team. There were 28 unique models submitted in the testing phase.

Table~\ref{tab:research-exp} summarizes the participants' past KBP experience and primary area of research. Interestingly,  $61.5\%$ of participants had no prior KBP experience and less than half ($42.6\%$) identified medical physics as their primary area of research. Machine learning researchers constituted the majority ($50.3\%$) of the participants, and only about one third ($33.3\%$) of those researchers had prior KBP experience. 

\begin{table}[H]
\caption{Participation throughout each phase of the Challenge.}
\setlength{\tabcolsep}{0.35em}	
\renewcommand{\arraystretch}{1}
\begin{tabular}{lccc}
\toprule
& Registration & Validation & Testing\\ \hline
Total participants & 195 & 73 & 54\\
Total teams & 129 & 44 & 28\\
Number of submissions & — & 1750 &28\\
\toprule
\end{tabular}
\label{tab:participant-info}
\end{table}

\begin{table}[H]
\caption{Distribution of participants by primary research area (rows) and whether they have past KBP research experience (columns).}
\setlength{\tabcolsep}{0.35em}	
\renewcommand{\arraystretch}{1}
\begin{tabular}{lcc|c}
\toprule
& \multicolumn{2}{c|}{\textbf{KBP Experience}}\\
\textbf{Primary Research Area} &   Yes &    No &  Total \\
\hline
Machine Learning      &  16.9\% &  33.3\% &   50.3\% \\
Medical Physics       &  17.9\% &  24.6\% &   42.6\% \\
Optimization          &   2.1\% &   2.6\% &    4.6\% \\
Other                 &   1.5\% &   1.0\% &    2.6\% \\
\hline
Total                 &  38.5\% &  61.5\% &  100.0\%\\
\toprule
\end{tabular}
\label{tab:research-exp}
\end{table}		

Table~\ref{tab:geo} presents the proportion of participants by country of work or study. In total, 28 different countries were represented in the Challenge. The three countries with the most participants were the United States (32.8\%), China (17.4\%), and India (11.3\%). Each of the other 25 counties that were represented had less than 5.1\% of the participants.  

\begin{table}[H]
\caption{The proportion of participants based on country of work or study.}
\setlength{\tabcolsep}{0.5em}	
\renewcommand{\arraystretch}{1}
\begin{tabular}{lc | lc | lc | lc}
\toprule
\centering
Australia	&	2.1\%	&	Colombia	&	4.1\%	&	Malaysia 	&	0.5\%	&	Sudan	&	0.5\%	\\
Austria	&	2.1\%	&	Croatia	&	0.5\%	&	Netherlands	&	1.0\%	&	Sweden	&	1.5\%	\\
Bangladesh	&	0.5\%	&	Finland	&	2.1\%	&	Pakistan	&	1.0\%	&	Taiwan	&	2.1\%	\\
Belgium	&	1.5\%	&	France	&	3.6\%	&	Poland	&	0.5\%	&	Turkey	&	1.0\%	\\
Brazil	&	0.5\%	&	Germany	&	1.0\%	&	South Africa	&	0.5\%	&	United Kingdom	&	1.0\%	\\  
Canada	&	5.1\%	&	India	&	11.3\%	&	South Korea	&	3.1\%	&	United States	&	32.8\%	\\
China	&	17.4\%	&	Japan	&	1.0\%	&	Spain	&	1.0\%	&	Vietnam	&	0.5\%	\\ 
\toprule
        \end{tabular}
\label{tab:geo}
\end{table}		

In Table~\ref{tab:EDI}, we present the aggregate data from our EDI survey. 
Men were overrepresented in OpenKBP (76.9\%) compared to the science and engineering population (52.3\%) and the general US population (49.2\%).  ``Asian American/Asian'' was the most common racial or ethnic identity (48.7\%) in OpenKBP, much greater than the science and engineering population (13.0\%) and the general US population (5.6\%). On the other hand, individuals who identified as ``White'' were underrepresented in OpenKBP (21.5\%) compared to the science and engineering (68.7\%) and US (60.0\%) populations. Individuals who identified as ``African American/Black'' (1.0\%) and ``Hispanic/Latinx'' (4.1\%) were also underrepresented relative to both baseline populations. A relatively large proportion (18.9\%) of respondents chose not identify their racial or ethnic identity. Lastly, the proportion of OpenKBP participants who identified as having no disability (87.2\%) was comparable to both the science and engineering (89.7\%) and general US (87.3\%) population. Fewer respondents identified with having a disability (2.1\%) compared to both baselines (10.3\% and 12.7\%), and the remaining proportion of OpenKBP participants chose not to identify their disability status.

\begin{table}[H]
\caption{The equity, diversity, and inclusion data (rows) of three populations of people (columns). In order, the columns correspond to the population people who participated in the Challenge (OpenKBP), are employed in the United States in science and engineering (S\&E), and live in the United States (US). A dash (\textemdash) indicates that the data is unavailable.}
\setlength{\tabcolsep}{0.35em}	
\renewcommand{\arraystretch}{1}
\begin{tabular}{llccc}
\toprule
& & OpenKBP & S\&E\cite{nsf_stats} & US\cite{us-population} \\
\multicolumn{2}{l}{\textbf{Number of people ($n$)}} & 195 & 27,274 & 328,239,523 \\
\multicolumn{2}{l}{\textbf{Gender identity}} \\
&Man                          &  76.9\% &  52.3\% & 49.2\% \\
&Woman                        &  12.8\% &  47.7\% & 50.8\%\\
&Prefer not to say            &   6.7\% & \textemdash & \textemdash \\
&No answer                    &   3.6\% &  \textemdash & \textemdash\\
\multicolumn{2}{l}{\textbf{Racial or ethnic identity}} \\
&African American/Black       &   1.0\% & 7.3\% & 12.4\%\\
&Asian American/Asian         &  48.7\% & 13.0\% &  5.6\%\\
&Hispanic/Latinx              &   4.1\% & 8.6\% & 18.4\% \\
&Middle Eastern/North African &   2.1\% & \textemdash & \textemdash\\
&Native American/Indigenous   &   0.5\% & 0.3\% & 0.7\%\\
&Native Hawaiian/Other Pacific Islander   &   0.0\% & 0.3\% & 0.2\%\\
&White                        &  21.5\% & 68.7\% & 60.0\%\\
&Other                        &   3.1\% & 1.8\% & 2.8\% \\
&Prefer not to say            &  13.3\% & \textemdash & \textemdash \\
&No answer                    &   5.6\% & \textemdash & \textemdash\\
\multicolumn{2}{l}{\textbf{Identify as having a disability}} \\
&No           &  87.2\% & 89.7\% & 87.3\%\\
&Yes            &   2.1\% & 10.3\% & 12.7\%\\
&Prefer not to say            &   7.2\% & \textemdash & \textemdash\\
&No answer                    &   3.6\% & \textemdash & \textemdash\\
\toprule
\end{tabular}
\label{tab:EDI}
\end{table}	



\subsubsection{Performance over validation phase}
\begin{figure}[H]
\centering
\subfigure[\ Dose score]{
\includegraphics[width=0.4675\linewidth]{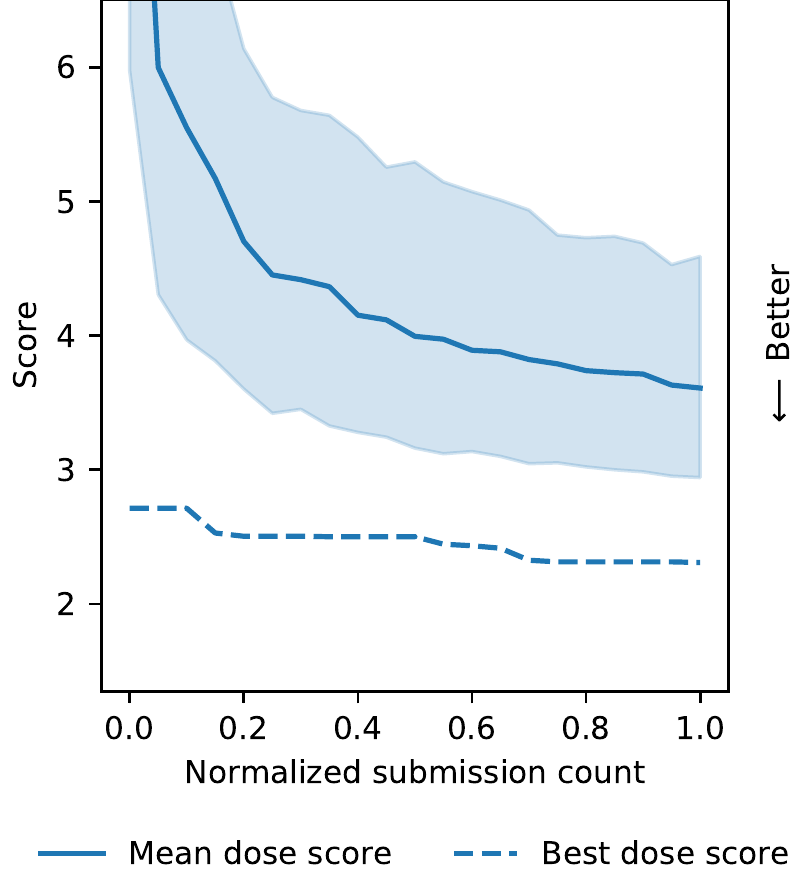} }
\subfigure[\ DVH score]{
\includegraphics[width=0.4675\linewidth]{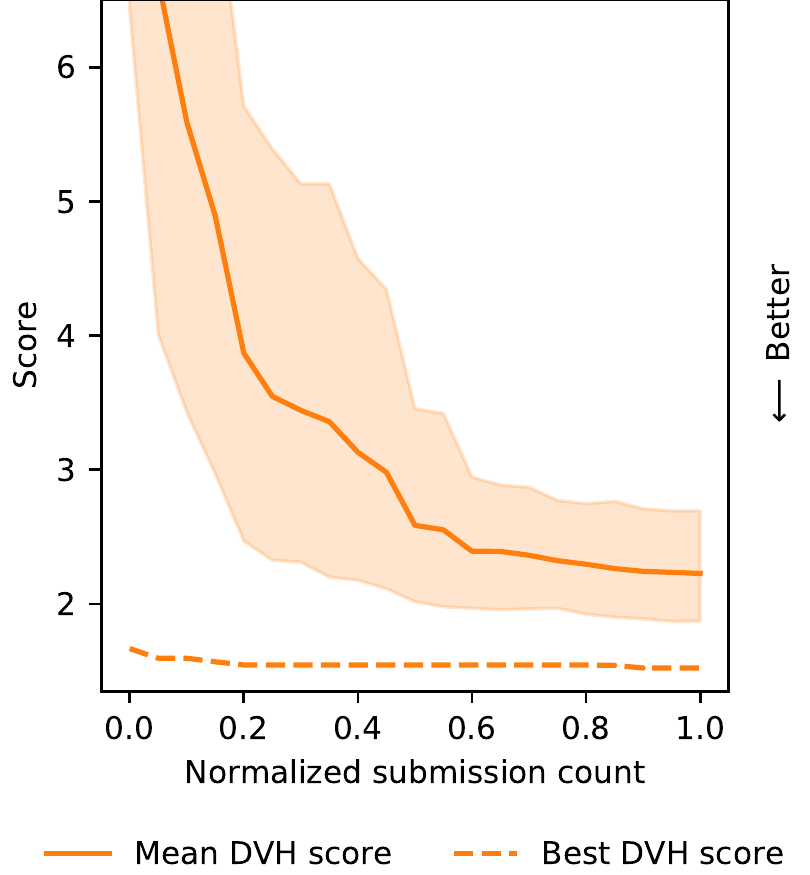} }
        \caption{The distribution of the dose and DVH scores across all teams. The solid lines indicate the mean score and the shaded regions indicate the 95\% confidence interval. A dash lined indicates the best score.}
        \label{fig:validation-progress}
\end{figure}

In Figure~\ref{fig:validation-progress}, we plot the distribution of team scores against normalized submission count. The plots show that teams generally improved their model throughout the validation phase, however, most teams made the largest improvements early on. Overall, the average team improved their dose and DVH score by a factor of 2.7 and 5.7, respectively, over the course of the validation phase. Over all of the NSC bins, the best dose and DVH scores were achieved by two and four different teams, respectively. There were a total of seven lead changes throughout the validation phase.

\subsubsection{Final results in testing phase}
\begin{figure}[H]
\centering
         \includegraphics[width=\linewidth]{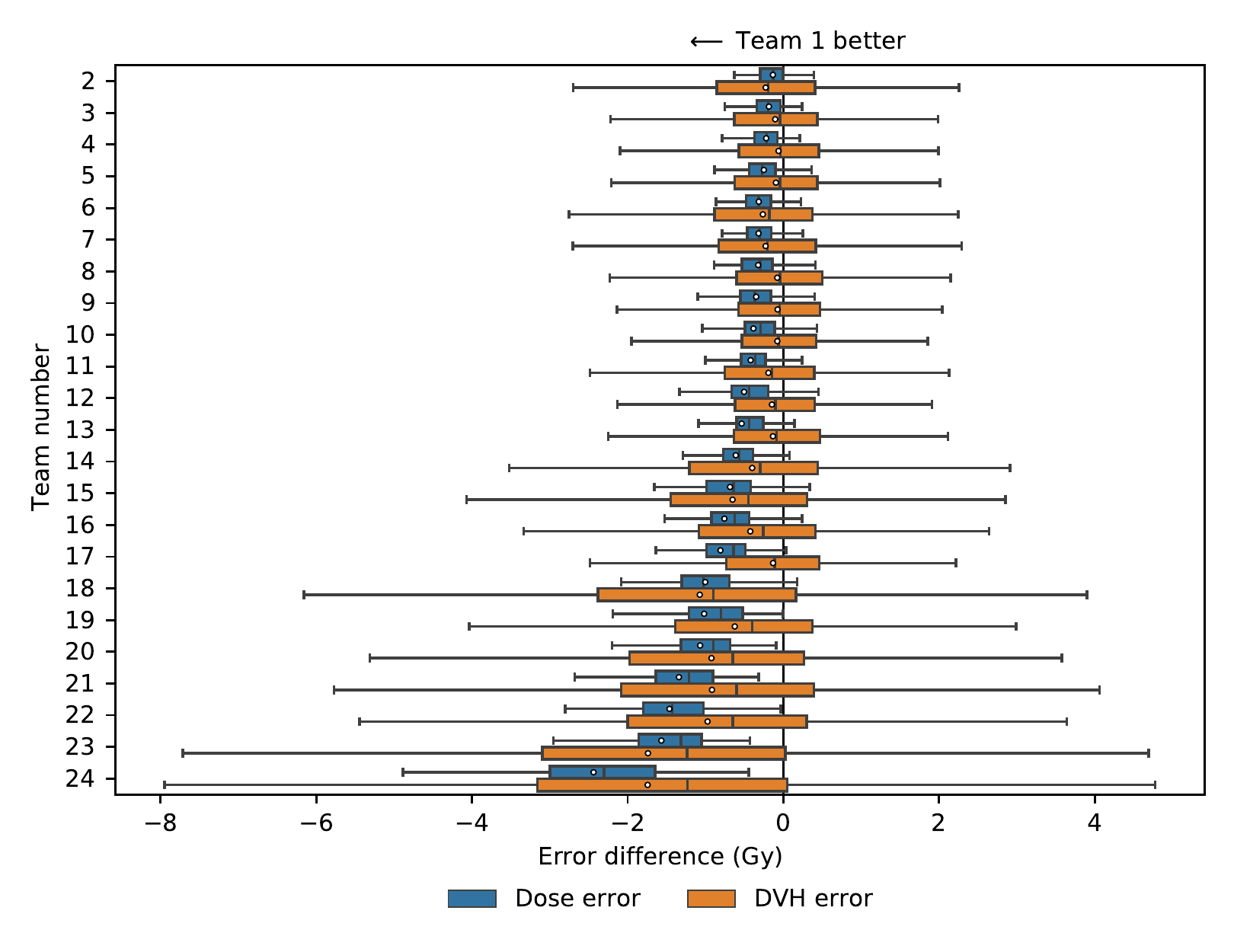}
        \caption{The distribution of dose and DVH error differences between the winning team (Team 1) and the top 23 runners-up, ranked by dose score. The boxes indicate median and IQR, and a circle indicates the mean. Whiskers extend to the minimum of 1.5 times the IQR and the most extreme outlier.} 
        \label{fig:test-scores}
\end{figure}

Figure~\ref{fig:test-scores} shows the distribution of error differences between the winning team (i.e., Team 1) and the top 23 runners-up (Teams 2-24). Compared to each of the other teams, Team 1 achieved significantly lower dose error over all 100 patients in the testing set ($P<0.05$) and significantly lower DVH error over all 1783 DVH criteria ($P<0.05$). Additionally, when compared to any other team, Team 1 achieved a lower dose and DVH error over at least 75\% and 52\% of patients and criteria, respectively. 

Table~\ref{tab:test-scores} summarizes the relative performance of each team under the MAE-based and MSE-based scores.  The winner and first runners-up according to MAE-based score would have finished in the same place under the MSE-based score. The average absolute rank difference between the two scoring approaches was one. The maximum rank difference was four and five for dose score and DVH score, respectively. The Spearman's rank-order correlation coefficient for the MAE-based and MSE-based score ranks was 0.983 ($P < 0.001$) and 0.981 ($P < 0.001$) for the dose and DVH score, respectively. Thus, the results of the competition would have been nearly identical had the MSE-based score been used.

\begin{table}[H]
\caption{The score and rank that each team achieved in the testing phase according to an MAE-based (i.e., the score used in the Challenge) and MSE-based (i.e., an alternative score) score. A positive difference in rank implies that a team performed better on the MSE-based score than on the MAE-based score.}
\setlength{\tabcolsep}{0.35em}	
\renewcommand{\arraystretch}{1}
\begin{tabular}{c || cc | cc | c || cc | cc | c}
\toprule
Team & \multicolumn{5}{c||}{Dose score} & \multicolumn{5}{c}{DVH score}\\
Number& \multicolumn{2}{c|}{MAE} & \multicolumn{2}{c|}{MSE} & Rank & \multicolumn{2}{c|}{MAE} & \multicolumn{2}{c|}{MSE} & Rank \\
& Score & Rank & Score & Rank & Difference & Score & Rank & Score & Rank & Difference \\
\hline
1  &      2.429 &          1 &         15.488 &              1 &                0 &     1.478 &         1 &         5.913 &             1 &               0 \\
2  &      2.564 &          2 &         16.550 &              2 &                0 &     1.704 &        12 &         6.812 &            10 &               2 \\
3  &      2.615 &          3 &         17.774 &              3 &                0 &     1.582 &         7 &         6.961 &            12 &              -5 \\
4  &      2.650 &          4 &         18.091 &              5 &               -1 &     1.539 &         2 &         6.031 &             2 &               0 \\
5  &      2.679 &          5 &         18.023 &              4 &                1 &     1.573 &         6 &         6.525 &             6 &               0 \\
6  &      2.745 &          6 &         19.213 &              7 &               -1 &     1.741 &        14 &         7.088 &            14 &               0 \\
7  &      2.748 &          7 &         18.916 &              6 &                1 &     1.706 &        13 &         7.083 &            13 &               0 \\
8  &      2.753 &          8 &         19.356 &              8 &                0 &     1.556 &         5 &         6.372 &             3 &               2 \\
9  &      2.778 &          9 &         19.469 &              9 &                0 &     1.551 &         3 &         6.634 &             7 &              -4 \\
10 &      2.814 &         10 &         23.417 &             13 &               -3 &     1.555 &         4 &         6.433 &             4 &               0 \\
11 &      2.850 &         11 &         20.432 &             10 &                1 &     1.669 &        11 &         6.904 &            11 &               0 \\
12 &      2.934 &         12 &         25.266 &             14 &               -2 &     1.624 &        10 &         6.699 &             8 &               2 \\
13 &      2.965 &         13 &         25.564 &             16 &               -3 &     1.611 &         9 &         6.745 &             9 &               0 \\
14 &      3.040 &         14 &         23.192 &             12 &                2 &     1.878 &        15 &         7.648 &            15 &               0 \\
15 &      3.114 &         15 &         22.454 &             11 &                4 &     2.130 &        18 &         9.445 &            17 &               1 \\
16 &      3.186 &         16 &         25.427 &             15 &                1 &     1.902 &        16 &         8.622 &            16 &               0 \\
17 &      3.237 &         17 &         33.642 &             20 &               -3 &     1.610 &         8 &         6.441 &             5 &               3 \\
18 &      3.432 &         18 &         27.046 &             17 &                1 &     2.551 &        22 &        12.602 &            22 &               0 \\
19 &      3.447 &         19 &         31.769 &             18 &                1 &     2.101 &        17 &        10.830 &            18 &              -1 \\
20 &      3.498 &         20 &         33.459 &             19 &                1 &     2.401 &        20 &        11.324 &            19 &               1 \\
21 &      3.771 &         21 &         37.980 &             22 &               -1 &     2.394 &        19 &        11.734 &            20 &              -1 \\
22 &      3.892 &         22 &         37.439 &             21 &                1 &     2.451 &        21 &        12.436 &            21 &               0 \\
23 &      3.996 &         23 &         43.654 &             23 &                0 &     3.216 &        23 &        19.856 &            23 &               0 \\
24 &      4.867 &         24 &         55.435 &             24 &                0 &     3.220 &        24 &        20.178 &            24 &               0 \\
25 &      5.446 &         25 &         70.694 &             25 &                0 &     3.712 &        25 &        22.335 &            25 &               0 \\
26 &      8.165 &         26 &        130.212 &             26 &                0 &    10.362 &        28 &       143.358 &            27 &               1 \\
27 &     11.818 &         27 &        266.484 &             27 &                0 &     7.713 &        26 &       100.080 &            26 &               0 \\
28 &     14.660 &         28 &        341.024 &             28 &                0 &     8.379 &        27 &       155.465 &            28 &              -1 \\
\toprule
\end{tabular}
\label{tab:test-scores}
\end{table}	

\subsubsection{Common modeling decisions}
According to the model survey, every team in the testing phase trained neural networks to predict dose distributions. The majority of those models had architectures based on U-Net,\cite{Unet} V-Net,\cite{Vnet} and Pix2Pix\cite{isola:2017image} models. All models were built using either a \textsf{TensorFlow} (Google AI, US) or \textsf{PyTorch} (Facebook AI Research, US) framework, but many teams also reported using higher-level libraries like \textsf{fast.ai}\cite{fastai} to simplify model development. To train and develop the models quickly, teams generally used a GPU (e.g., NVIDIA 1080Ti, NVIDIA Titan V); seven teams also reported that they used \textsf{Google Colab}.

Many teams used generalizable techniques to get better model performance. For example, 22 of the 28 teams used some form of data augmentation in their training process, and 15 teams combined two or more augmentation methods. Common forms of data augmentation were rotations, flips, crops, and translations. Most teams also reported that they normalized dose and CT Hounsfield units. Additionally, most teams used a standard loss function, e.g., MAE, MSE, GAN loss. There were also some teams that developed radiation therapy specific loss functions (e.g., functions that prioritized regions-of-interest more than the unclassified tissue). Lastly, ensemble methods were used by several of the top teams. Those methods used multiple neural networks to predict candidate dose distributions that were combined by taking the average prediction.

\section{Discussion}
%

There is widespread research interest in knowledge-based planning (KBP) dose prediction methods. However, the lack of standardized metrics and datasets make it difficult to measure progress in the field.  In this paper, we present the first set of standardized metrics and the first open-access dataset for KBP research as part of the OpenKBP Grand Challenge, the first competition for KBP research. The Challenge democratizes KBP research by enabling researchers without access to clinical radiation therapy plans to develop state-of-the-art dose prediction methods. This spurred the development of 28 unique models and will serve as an important benchmark as field of KBP  continues to grow. 

Our open-access dataset contains real patient images that were contoured by clinicians at twelve institutions with different planning protocols. There are two major differences in protocol that introduce some variance in how PTVs were drawn. First, the raw public clinical data included plans with multiple radiation therapy modalities. For example, some of the institutions delivered hybrid-IMRT/3DCRT plans, and those plans had no PTV margins on the lower neck target volumes. Second, some of the raw public clinical data is from multiple trials with unconventional contouring in the extent of the target volumes. For example, we observed some anisotropic PTV margins that were clipped to omit the OARs. \ab{These variations are non-existent in the raw private clinical dataset, which contains plans from a single institution where all planning and contouring was done according to a standard process, that was used to create the dose distributions for the competition.} This variation may have been a factor in the public synthetic dose being non-inferior to the private clinical dose on 19 of 23 dose-volume criteria.

We proposed two new metrics that quantify the general performance (i.e., dose score) and the clinical performance (i.e., DVH score) of dose prediction methods. These two metrics may help measure progress in KBP research, and they will complement other metrics that are typically used in the literature to quantify strengths and weaknesses of a model. Other metrics are still important because our scoring metrics are unable to quantify every facet of radiation therapy dose quality. For example, the DVH criteria evaluated for the DVH score have varying degrees of clinical importance (e.g., $\D{max}{\text{mandible}}$ is much more important than $\D{mean}{\text{mandible}}$). We chose to weigh all errors equally because quantifying relative clinical importance is non-trivial and largely dependent on the institution. Additionally, since the scores are unweighted it is straightforward to use the scores for all other sites that have OARs and targets (e.g., prostate).

We aimed to make OpenKBP as accessible as possible in order to build a large and inclusive community. \ab{By building a large and inclusive community we can ensure that underrepresented populations can contribute to KBP research, which should both accelerate innovation\cite{Hofstra:2020aa} and improve the quality of healthcare.}\cite{Gomez:2019aa} 
As part of this Challenge, we released all competition data in a non-proprietary format (comma-separated value) and a well-documented code repository that helped participants use the data easily and efficiently in \textsf{Python} without costly commercial software. The code repository also had instructions to give all participants access to high-quality computational resources at zero cost (i.e., \textsf{Google Colab}). In an effort to keep the data manageable for all participants, we also opted to use relatively large voxels (e.g., $3\text{mm}\times3\text{mm}\times2\text{mm}$ voxels), which ensured that the dose prediction problem was tractable for anyone using \textsf{Google Colab}. We conjecture that this manageable data size also helped the teams iterate and improve their models, which is reflected by the number of submissions made by teams in the validation phase (40 submissions on average).

A limitation of this work is that is uses synthetic dose distributions to augment the real clinical data. Those dose distributions were generated by a published KBP pipeline\cite{Babier:2018b} and filtered via Algorithm~\ref{algo:filter}, however, they underwent less scrutiny than clinical plans. Extensions of this work should ensure that the top performing models on this dataset also perform well with clinical dose distributions. A second limitation is that we can only report commonalities between the top models, which are correlated attributes rather than causal attributes. Future work should do ablation testing to isolate exactly what attributes contribute to a good dose prediction model. Lastly, all dose predictions were evaluated and ranked based on two scores. These scores do not capture all of the strengths and weaknesses of the models submitted to the Challenge.

\section{Conclusion}
OpenKBP democratizes knowledge-based planning research by making it accessible to everyone. It is also the first platform that researchers can use to compare their KBP dose prediction methods in a standardized way. The Challenge helps validate our platform and provides a much needed benchmark for the field. This new platform should help accelerate the progress in the field of KBP research, much like how ImageNet helped accelerate the progress in the field of computer vision. 

\section{Acknowledgments}
OpenKBP was supported by the American Association of Physicists in Medicine and its Working Group on Grand Challenges. The authors also gratefully acknowledge everybody who participated in the Challenge.

\bibliography{refs}

\appendix

\section{Data Format}
The data for OpenKBP is structured to facilitate the development and validation of dose prediction models. In this section, we describe how the data is stored and formatted.

\subsection{Summary of data}
The data for each patient is provided as comma-separated values (CSV) files, which are separated into directories with the corresponding patient number. The files for each patient include:

\begin{description}
\item [dose.csv] the full 3D dose distribution that was used to treat the patient (in units of Gy).
\item [ct.csv] grey-scale images of the patient prior to treatment (in Hounsfield units). There is a mix of 12-bit and 16-bit formats, and we recommend clipping the CT values to be between 0 and 4095 (inclusive) to convert them all to 12-bit number formats, which is the more common convention. 
\item [voxels.csv] The dimensions of the patient voxels (in units of mm).
\item [possible\_dose\_mask.csv] a mask of voxels that can receive dose (i.e., the dose will always be zero where this mask is zero). 
\item [Structure masks] a mask that labels any voxel that is contained in the respective structure. The tensor for each structure stored as a CSV file under its respective structure name. Only structures that were contoured in the patient have CSV files.  
\begin{description}
\item [Brainstem.csv] mask of brainstem voxels.
\item [SpinalCord.csv] mask of spinal cord voxels.
\item [RightParotid.csv] mask of right parotid voxels.
\item [LeftParotid.csv] mask of left parotid voxels.
\item [Esophagus.csv] mask of esophagus voxels.
\item [Larynx.csv] mask of larynx voxels.
\item [Mandible.csv] mask of mandible voxels.
\item [PTV56.csv] mask of PTV56 voxels.
\item [PTV63.csv] mask of PTV63 voxels.
\item [PTV70.csv] mask of PTV70 voxels.
\end{description}
\end{description}

\subsection{Data format}
Other than the file voxels.csv, which contains a list of only three numbers, all of the CSV data in OpenKBP is saved as sparse tensors (i.e., only non-zero values are stored). The advantage of this sparse format, compared to dense tensors (i.e., all values are stored), is that the data size is smaller and thus loads into memory faster, which leads to faster model training. The disadvantage, is that working with sparse tensors is less intuitive than working with dense tensors. In general, we recommend converting the data into dense tensors once it is loaded. 

All of the sparse tensors are stored in CSV files with two columns. The first column contains a list of indices. The second column contains either a list of values for the corresponding indices, or it contains no values if the tensor is a mask (i.e., where all corresponding values are 1). All indices are stored as single numbers that unravel into a 3D (i.e., x-y-z) coordinate system via C-contiguous ordering. We provide \textsf{Python} code in our repository to load the data as dense tensors.

\newpage 

\section{Surveys}
In this section, we present the two mandatory surveys that we released during the Challenge. In each survey, respondents answered questions either by writing free-text or by selecting option(s) from a list. 

Mandatory questions are marked with an asterisk (*).

\subsection{Registration}
All participants completed the following two part survey to register for OpenKBP. 

\renewcommand{\descriptionlabel}[1]{%
  \hspace\labelsep \upshape\bfseries #1\ %
}
\newcommand{\surveyAlign[1]}{\parbox[#1]{\linewidth}}
\subsubsection{Part 1: Professional information}
Please complete this form to be given access to the OpenKBP competition. \\

\renewcommand\labelitemi{${\mathrel{\bigcirc}}$}
\def\surveyQ#1{\noindent\textbf{#1}\\\vspace{-0.75cm}}
\def\surveyA#1{\hspace{1cm}{#1}\\\vspace{0.5cm}}
\def\surveyAMC{
  \begin{itemize}
  \setlength\itemsep{0em}
  \vspace{0.5em}
  }

\surveyQ{First Name*}
\surveyA{Short-answer text}
    
\surveyQ{Last Name*}
\surveyA{Short-answer text}

\surveyQ{E-mail (must be the same address used for your CodaLab account)*}
\surveyA{Short-answer text}
    
\surveyQ{Institution name without acronyms (University, Hospital, Company, etc.)*}
\surveyA{Short-answer text}

\surveyQ{Department (Computer Science, Medical Biophysics, Radiation Oncology, Machine Learning, Industrial Engineering, etc.)*}
\surveyA{Short-answer text}

\surveyQ{Primary research area*}
\surveyAMC
\item Medical Physics
\item Machine Learning
\item Optimization
\item Other...
\end{itemize}

\surveyQ{Have you done research in knowledge-based planning in the past?*}
\surveyAMC
\item Yes
\item No
\end{itemize}

\surveyQ{Position*}
\surveyAMC
\item Student
\item Professor
\item Post doctoral fellow
\item Medical physicist
\item Radiation oncologist
\item Industry research
\item Other...
\end{itemize}

\subsubsection{Part 2: Equity, diversity, and inclusion}
To help us learn how to support the diversification of researchers in the OpenKBP Grand Challenge, we ask that all applicants complete an equity survey at the time of their registration. Equity is one of our competition’s goals. We seek to remove barriers to participation for all people including women, LGBTQ individuals, persons with disabilities, Indigenous People and racialized persons and persons of colour.

Your participation is voluntary and your responses are confidential. We hope you will choose to answer these questions to help us bring you an even better competition next time. The information we receive from your responses will be used to better understand who has access to the competition, to identify barriers that may exist and areas to develop and/or improve in our rules and procedure to achieve more diversity and equity in the application process. All responses will be kept strictly confidential and will be reported only in aggregate so that you cannot be personally identified by your characteristics.\\

\renewcommand\labelitemi{$\Box$}
\surveyQ{Do you self-identify as (choose all that apply):}
\surveyAMC
    \item Man
    \item Women
    \item Transgender
    \item Prefer not to say
    \item Other...
\end{itemize}

\surveyQ{Please indicate the racial or ethnic groups with which you identify (check all that apply):}
\surveyAMC
\item African American/Black
\item Asian American/Asian
\item Hispanic/Latinx
\item Middle Eastern/North African
\item Native American/Indigenous
\item Native Hawaiian/Other Pacific Islander
\item White
\item Prefer not to say
\item Other...
\end{itemize}

\renewcommand\labelitemi{$\bigcirc$}
\surveyQ{Do you identify as a person with a disability? This may mean that either you: (i) have a long-term or recurring condition or health problem which limits the kind or amount of work you can do in the workplace; OR (ii) feel that you may be perceived as limited in the kind or amount of work which you can do because of a physical, mental, sensory, psychiatric, or learning impairment.}
\surveyAMC
    \item Yes, I identify as a person with a disability
    \item No, I do not identify as a person with a disability
    \item Prefer not to say
\end{itemize}

\subsection{Model summary}
Every team that competed in the testing phase of the Challenge also completed the following one part survey to summarize their model. 

\subsubsection{Model survey}
Please describe your final dose prediction model using this survey. We will consider your submission complete only if this survey is submitted. Any submission made on CodaLab that is not associated with a survey response will be considered void, and it will not be ranked in the final leaderboard. We may also reach out to you for more information.

This survey includes 5 long answer questions, and we expect the cumulative word count of your responses to be about 350 words. We provide an estimated word count for your response to each question. These estimates are only a guide and you may provide more detail where you see fit. Please reach out if you have any questions or need clarification. 

For teams, only one team member should submit this form. \\

\renewcommand\labelitemi{$\bigcirc$}
\surveyQ{Username on CodaLab*}
\surveyA{Short-answer text}
    
\surveyQ{Team Name on CodaLab (enter N/A if you have no team)*}
\surveyA{Short-answer text}
    
\surveyQ{Broadly speaking, how would you describe your model?*}
\surveyAMC
\item Linear regression
\item Random forest
\item Neural network
\item Gradient boosted trees
\item Support vector machines
\item Other...
\end{itemize}

\surveyQ{Briefly describe your approach. ($\mathbf{\sim150}$ words)*}
\surveyA{Long-answer text}
    
\surveyQ{What would you say is the biggest contributing factor(s) to your models efficacy?  ($\mathbf{\sim100}$ words)*}
\surveyA{Long-answer text}
    
\surveyQ{Did you use transfer learning?*}
\surveyAMC
\item Yes
\item No
\end{itemize}

\surveyQ{Please describe any data augmentation methods that you used (e.g., rotations, private clinical dataset)? ($\mathbf{\sim50}$ words)*}
\surveyA{Long-answer text}
    
\surveyQ{Briefly describe your loss function. Did you use radiation therapy specific metrics in your loss function (e.g., max dose to PTV)? ($\mathbf{\sim50}$ words)*}
\surveyA{Long-answer text}
    
\surveyQ{Briefly describe the hardware (e.g., GPU model, CPU model) or cloud resources (e.g., Google Colab) that you used. ($\mathbf{\sim25}$ words)*}
\surveyA{Long-answer text}
    
\surveyQ{Please leave any other comments about your process here.}
\surveyA{Long-answer text}

\surveyQ{Provide a link to the code repository that will recreate your model. We will include links to all the provided repositories from our existing OpenKBP Github to enable new users to build on a library of existing models. You may also provide a repository link at a later date, but it is not required.}
\surveyA{Short-answer text}

\end{document}